# Montage based 3D Medical Image Retrieval from Traumatic Brain Injury Cohort using Deep Convolutional Neural Network

Cailey I. Kerley [1], Yuankai Huo [1]*, Shikha Chaganti [2], Shunxing Bao [2], Mayur B. Patel [3], Bennett A. Landman [1,2,4,5]

[1] Department of Electrical Engineering, Vanderbilt University, Nashville, TN, USA
[2] Department of Computer Science, Vanderbilt University, Nashville, TN, USA
[3] Departments of Surgery, Neurosurgery, Hearing & Speech Sciences; Center for Health Services Research, Vanderbilt Brain Institute; Critical Illness, Brain Dysfunction, and Survivorship Center, Vanderbilt University Medical Center; VA Tennessee Valley Healthcare System, Department of Veterans Affairs Medical Center, Nashville, TN, USA
[4] Departments of Radiology and Radiological Sciences, Vanderbilt University Medical Center, Nashville, TN, USA
[5] Department of Biomedical Engineering, Vanderbilt University, Nashville, TN, USA
[6] Institute of Imaging Science, Vanderbilt University, Nashville, TN, USA
(* Corresponding author: yuankai.huo@vanderbilt.edu)

**ABSTRACT**

Brain imaging analysis on clinically acquired computed tomography (CT) is essential for the diagnosis, risk prediction of progression, and treatment of the structural phenotypes of traumatic brain injury (TBI). However, in real clinical imaging scenarios, entire body CT images (e.g., neck, abdomen, chest, pelvis) are typically captured along with whole brain CT scans. For instance, in a typical sample of clinical TBI imaging cohort, only ~15% of CT scans actually contain whole brain CT images suitable for volumetric brain analyses; the remaining are partial brain or non-brain images. Therefore, a manual image retrieval process is typically required to isolate the whole brain CT scans from the entire cohort. However, the manual image retrieval is time and resource consuming and even more difficult for the larger cohorts. To alleviate the manual efforts, in this paper we propose an automated 3D medical image retrieval pipeline, called deep montage-based image retrieval (dMIR), which performs classification on 2D montage images via a deep convolutional neural network. The novelty of the proposed method for image processing is to characterize the medical image retrieval task based on the montage images. In a cohort of 2000 clinically acquired TBI scans, 794 scans were used as training data, 206 scans were used as validation data, and the remaining 1000 scans were used as testing data. The proposed achieved accuracy=1.0, recall=1.0, precision=1.0, f1=1.0 for validation data, while achieved accuracy=0.988, recall=0.962, precision=0.962, f1=0.962 for testing data. Thus, the proposed dMIR is able to perform accurate CT whole brain image retrieval from large-scale clinical cohorts.

**Keywords:** image retrieval, computed tomography, TBI, CT, traumatic brain injury, deep learning

## 1. INTRODUCTION

Traumatic brain injuries (TBI) affects more than 2 million patients every year in the United States, which may lead to severe physical, cognitive, and psychological disorders. Computed tomography (CT) on the human brain is one of the most widely used tools for diagnosing TBI associated with intracranial hemorrhage [1, 2]. In recent years, large-scale informatics analyses have shown its advantages in understanding TBI [3]. Thus, it is appealing to perform large-scale imaging analyses on TBI cohorts as well, especially as brain trauma is a heterogenous condition with large inter-subject variations. The challenge even in a single medical center or healthcare system is the generation of more than thousands of whole brain CT scans/year for TBI patients associated with a 5-fold higher number of non-brain scans (e.g., neck, abdomen, chest, pelvis). Therefore, manual image retrieval procedures are commonly included to filter out the unrelated scans. For instance, in a sample of clinically acquired TBI cohort at Vanderbilt University Medical Center (VUMC), only 307 scans (15%) are marked as useable whole brain CT scans in a set of 2000 scans. Moreover, the manual image retrieval is resource and time consuming on 2000 scans and could be impractical for larger cohorts. Previous efforts have been made on automatic medical image retrieval [4]. In recent years, the deep convolutional neural network (DCNN) based methods have been used in medical image retrieval due to the superior performance in speed and accuracy [5, 6]. However, most previous

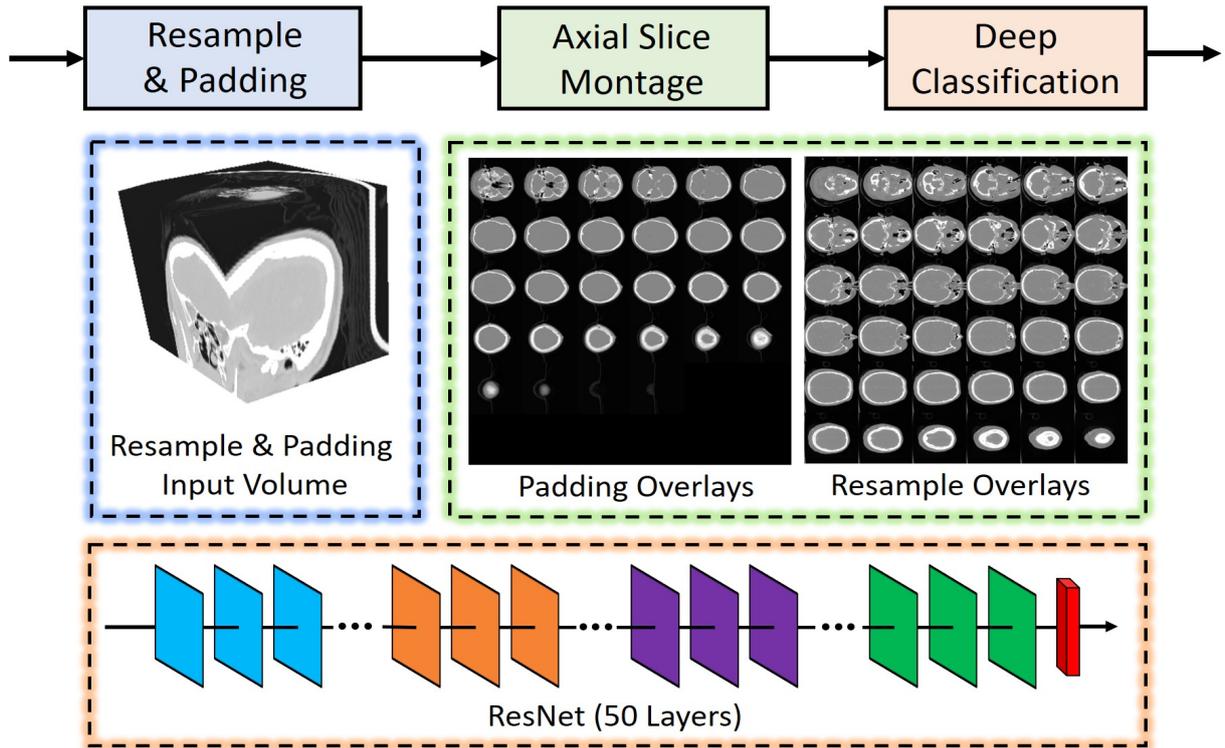

Figure 1. The dMIR whole brain CT scan retrieval method is presented. The 3D CT scans were resampled to in-plane resolution 512 × 512. If the slice number $z$ is larger than 36, the volume is resampled to 512 × 512 × 36. Otherwise, the 3D volume is resampled to 512 × 512 × $z$ and zero padding is performed to achieve a 512 × 512 × 36 volume. Next, the 512 × 512 × 36 volume was used to form the 2D montage figures. Finally, the 2D montage figures were used to train and test a deep neural network using a 50 layer ResNet.

DCNN medical image retrieval methods used only a single 2D slice without fully using the 3D information. Therefore, image retrieval methods for 3D medical images are desired for the community, especially for big medical image data processing [7].

In this paper, we propose deep montage-based image retrieval (dMIR) method, an automated 3D medical image retrieval pipeline, to retrieve usable whole brain CT scans from a heterogenous clinical TBI cohort. The 2D montage images from 3D scans are used as the training and testing images for training a classification deep convolutional neural network. ResNet [8] was employed as the DCNN network due to its superior performance in 2D image classification.

## 2. METHODS

### 2.1 Montage Image Generation

The framework of dMIR pipeline is shown in Figure 1, whose raw input is a 3D CT scan. Rather than learning from the 3D volume directly, we used 2D montage images to represent such 3D volumes. The original size of the input scan $S$ is $x \times y \times z$, which is then resampled to volume $V$ with 512 × 512 × 36 if $z \geq 36$ using the following steps

$$N = \text{floor}(\frac{z}{36}) \qquad (1)$$

$$m = \text{floor}\left(\frac{z}{2}\right) - N \times 17 \qquad (2)$$

$$V(:,:,i) = S(:,:,m + N \times i), \quad i \in [1,2,\ldots,36] \qquad (3)$$

where $N$ is the length of sampling step, $m$ is the first slice in $S$ to perform sampling. Otherwise, if $z < 36$, the scan is first resampled to 512 × 512 × $z$, then zero padding to 512 × 512 × 36. After achieving the resampled 3D volume $V$, the 2D montage images are generated by concatenating all slices of 3D volume $V$ on z direction to a 6 × 6 image matrix.

## 2.2 DCNN based Image Retrieval

The input 2D montage images are the input of the DCNN, so the image retrieval becomes a 2D image classification problem, where class = 0 means unusable scans and class = 1 means usable whole brain CT scans. The ResNet has been regarded as the de facto standard image classification network due to its superior performance in accuracy and training convergence. Therefore, the 50 layer ResNet (ResNet-50) was employed as the classification network in this study. The ResNet-50 network was trained from scratch with the number of input channel = 1 and the number of output channel = 2. All the montage input images are resampled to 512 × 512 for training purposes. More specifically, the maximum epoch = 100, batch size = 8, and learning rate = 0.0001. The Adam optimization algorithm [9] was used to train the network with weighted cross entropy loss. Since the classes are unbalanced, weights = [1, 10] are used as the weights for cross entropy loss. The experiment was performed on NVIDIA GForce Titian GPU with 12 GB memory. The DCNN was implemented in PyTorch 0.4 version [10].

## 3. DATA

Clinically acquired CT scans (n=2000) from VUMC were retrieved in deidentified form under Institutional review board approval and employed as training, validation, and testing data. Specifically, 794 scans from 80 subjects were used as training data; 206 scans from 21 subjects were used as validation data; 1000 scans from 551 subjects were used as testing data. In the 794 training scans, 124 are usable whole brain scans, while the remaining 670 are unusable scans. Among 206 validation scans, 27 are usable whole brain scans, while the remaining 179 are unusable scans. In 1000 testing scans, 156 are usable whole brain scans, while the remaining 844 scans are unusable scans. The content labels are performed by two human raters with visually checking on the montage figures. Figure 2 shows the montage figures for the usable whole brain CT scans. Figure 3 shows the montage figures for the first category of unusable scans, where the brains are not included in the scans. Figure 4 shows another category of unusable scans, which contained part of the brains but were dedicated for other head and neck structures. Those are marked as unusable since such scans were not preferred in the following brain image analyses, which typically required whole brain volumes on CT.

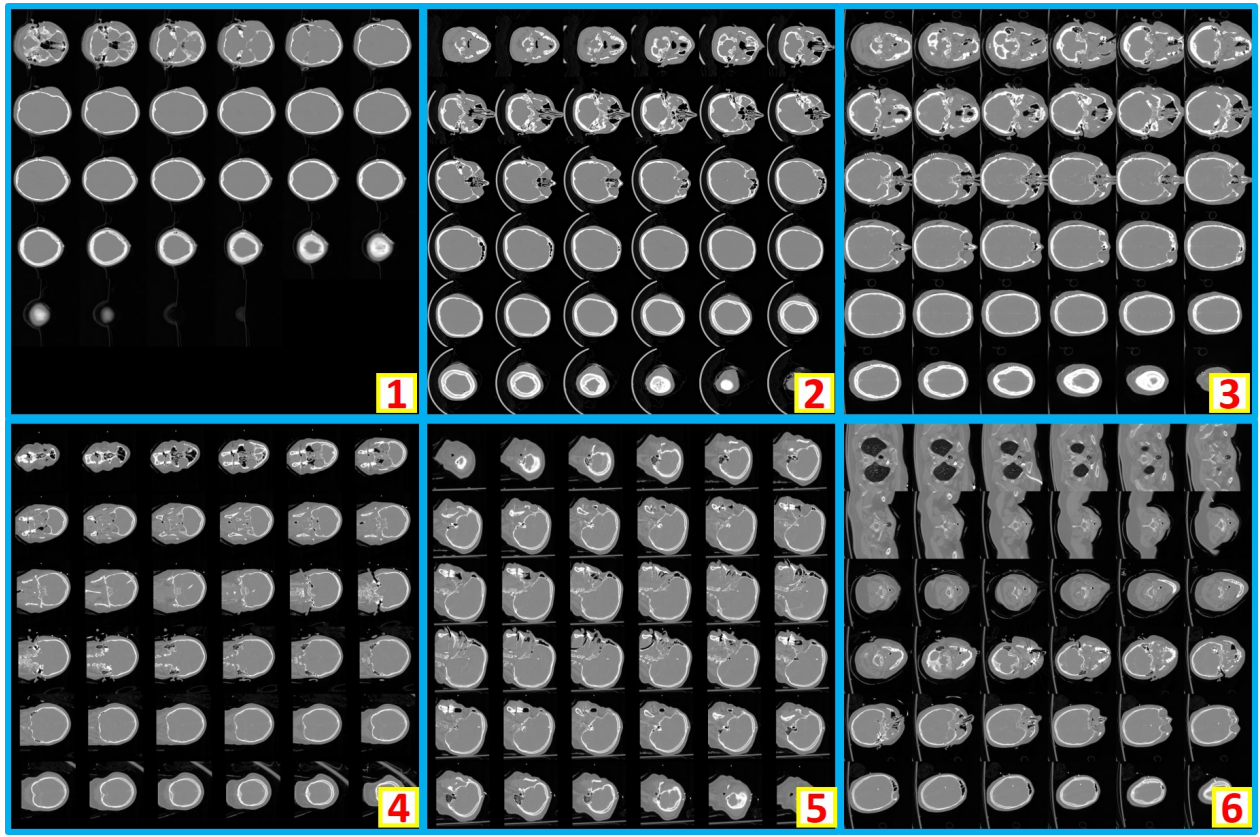

Figure 2. The example montage images of usable whole brain 3D CT scans. "1" shows a montage image with < 36 slices after zero padding. "2" and "3" are the montage images resampled from axial view images, while "4" and "5" are from coronal and saggital view images respectively. "6" presents an example that whole brain and part of body are included in one scan. As the entire brain volume is included, "6" is also regarded as usable.

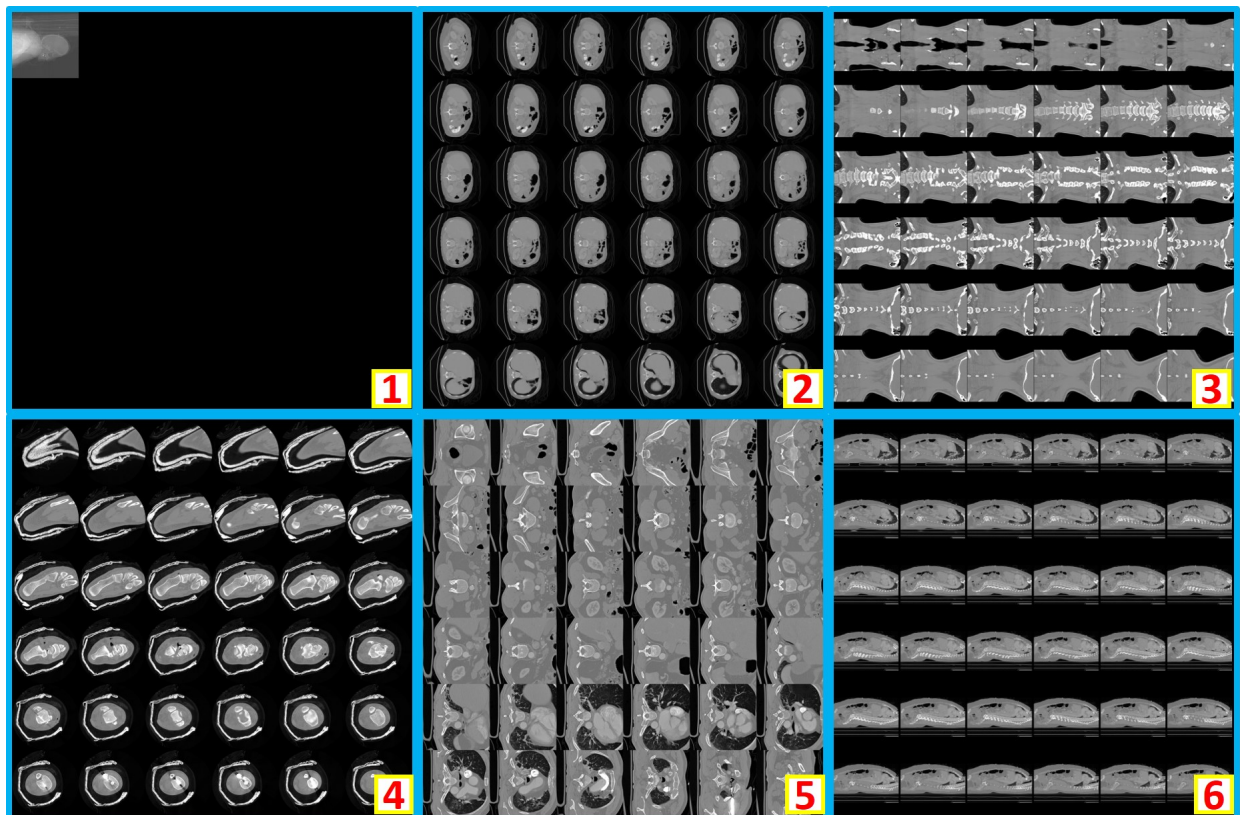

Figure 3. This figure shows six representative examples of unusable scans, which does not include brain in the scans.

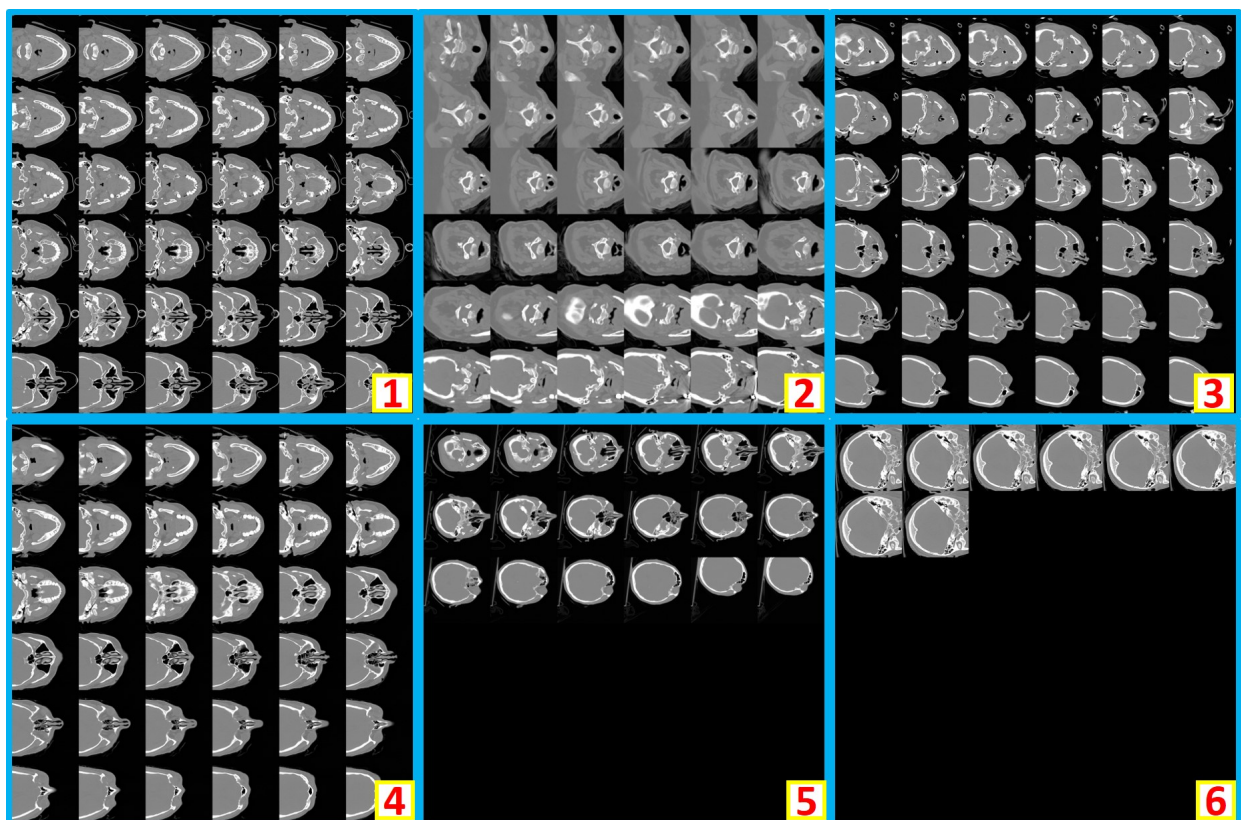

Figure 4. This figure shows six representative examples of unusable scans. "1" to "4" shows the incomplete brains (e.g., maxillofacial CT, neck CT, temporal bone CT, patient movement). In "5", an abnormal spatial translation is found in last two slices. For "6", the scan only covers few slices of the cranium.

## 4. RESULTS

The validation cohort was used to define the hyper-parameters in "Method" section. From the epoch number 42, the best validation performance was achieved on validation scans. From the empirical validation, the accuracy=1.0, recall=1.0, precision=1.0, f1=1.0 are achieved for 206 validation scans. The corresponding Receiver Operating Characteristic (ROC) curves and Area under the ROC Curve (AUC) are presented in Figure 5a. Then, the model at epoch number 42 was applied on the 1000 independent testing scans, which achieved accuracy=0.988, recall=0.962, precision=0.962, f1=0.962. The corresponding ROC curves and AUC are presented in Figure 5b. In total, 12 wrong classification images are obtained in the testing stage, where six are false positive and remaining six are false negative. To further exam the failures in Figure 5b, we plot all six false positive examples in Figure 6, while plots all six false negative examples in Figure 7.

## 5. DISCUSSION

In this paper, we presented the dMIR pipeline to perform usable whole brain CT image retrieval from heterogenous clinically acquired TBI scans. The proposed method achieved AUC=1 on validation cohort and AUC=0.9965 on independent validation cohort. The results show that the dMIR method is able to perform image retrieval in high accuracy. Thus, it is able to alleviate the heavy manual efforts for the image retrieval from clinical radiology archive. In this work, the ResNet-50 was employed as the classification network. However, we do not claim the ResNet-50 is the best network for this scenario. Since we have already achieved 1.0 accuracy on training and validation cohort, we did not compare the extra experiments on other deep image classification neural networks. It would be interesting to see if the performance of other networks (including ResNet) on larger training and testing cohorts.

The proposed method simplified the 3D whole volume learning problem to 2D learning task using the montage figures. The primary contribution of the proposed method for image processing is to provide a new way (montage figures) of performing medical image retrieval. As a result, the prevalent 2D based image retrieval methods developed in computer vision communities are able to be applied to medical images retrieval directly using the dMIR pipeline. In the future, the proposed method could be used to solve the quality assurance for large-scale image archives, such as Vanderbilt VUIIS-CCI XNAT [11].

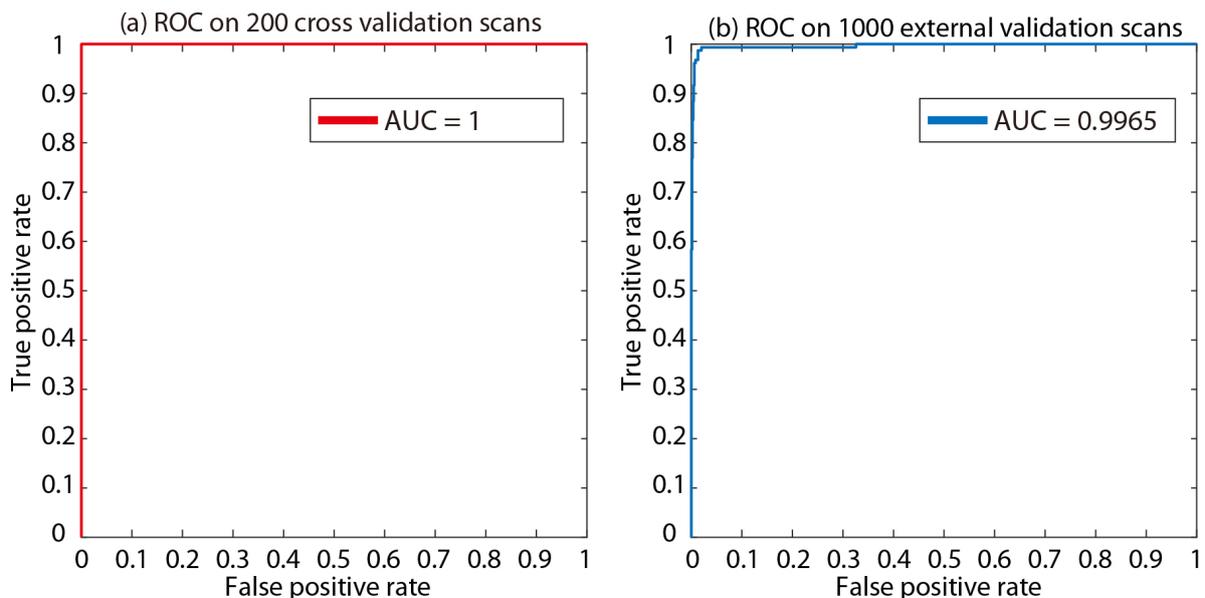

Figure 5. The Receiver Operating Characteristic (ROC) curves and Area under the ROC Curve (AUC) are presented. (a) is the ROC curve and AUC for 206 internal validation scans at the best performance epoch. After applying the model at such epoch on 1000 independent test scans, the AUC=0.9965 is achieved in (b).

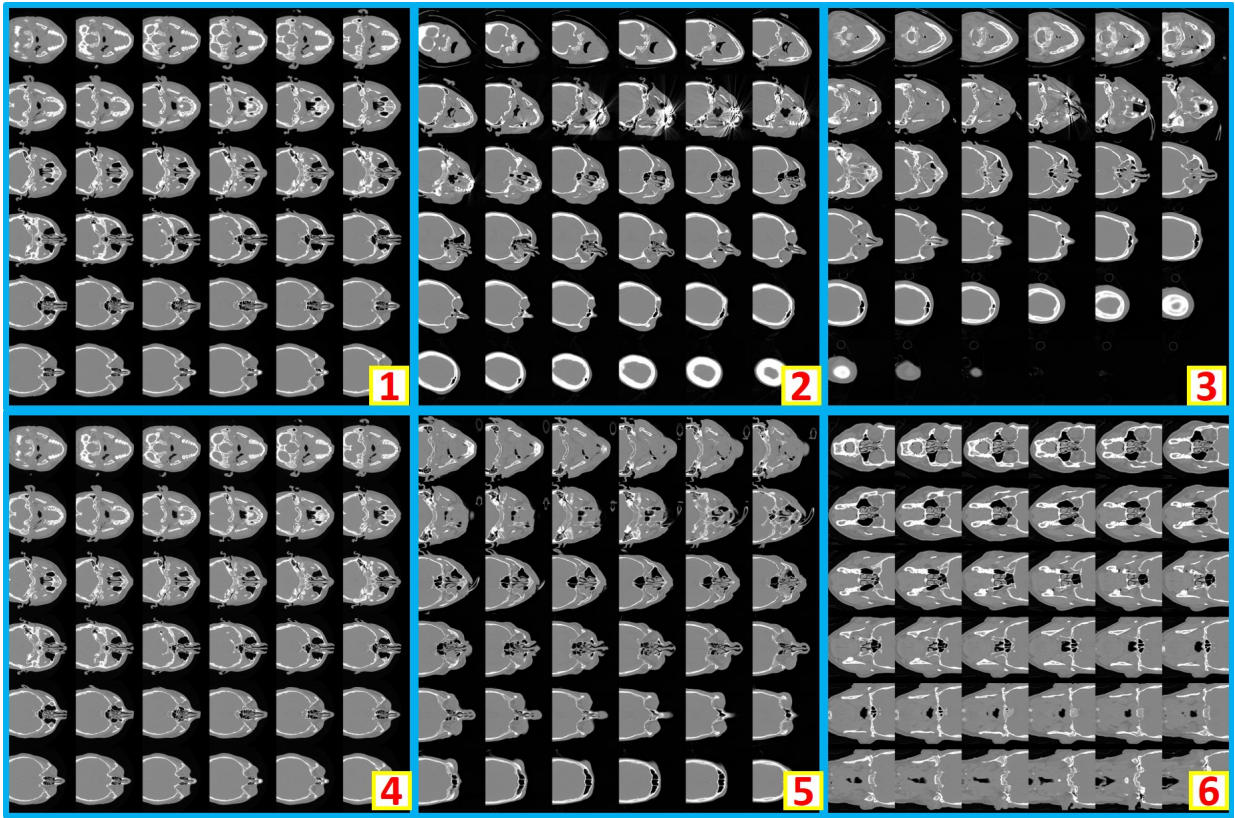

Figure 6. This figure shows all six false positive cases in testing stage. The false positive scans are typically partially incomplete whole brain CT scans.

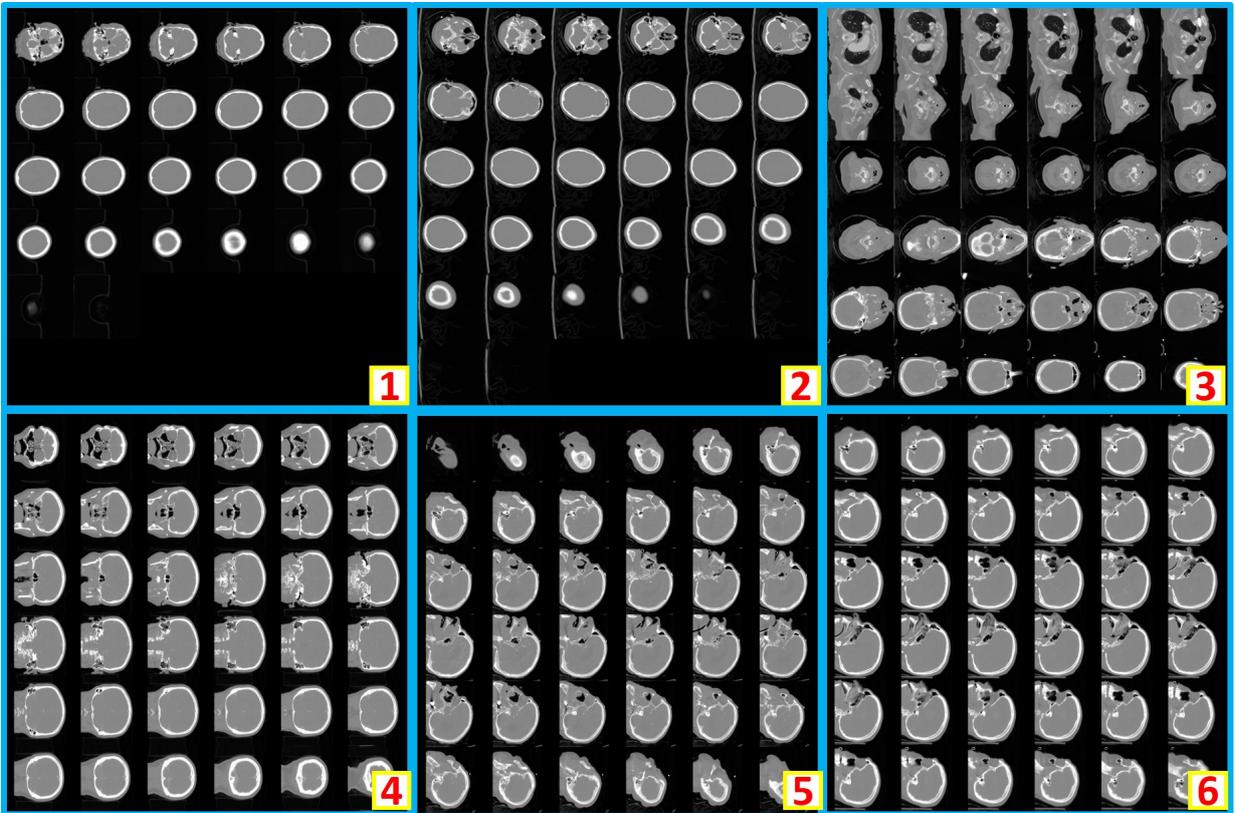

Figure 7. This figure shows all six false negative cases in testing stage.

## ACKNOWLEDGEMENT

This research was supported by NSF CAREER 1452485, NIH grants R03EB012461 (Landman), and R01GM120484 (Patel). This research was conducted with the support from Intramural Research Program, National Institute on Aging, NIH. This study was in part using the resources of the Advanced Computing Center for Research and Education (ACCRE) at Vanderbilt University, Nashville, TN. This project was supported in part by ViSE/VICTR VR3029 and the National Center for Research Resources, Grant UL1 RR024975-01, and is now at the National Center for Advancing Translational Sciences, Grant 2 UL1 TR000445-06. We gratefully acknowledge the support of NVIDIA Corporation with the donation of the Titan X Pascal GPU used for this research. The imaging dataset(s) used for the analysis described were obtained from ImageVU, a research resource supported by the VICTR CTSA award (ULTR000445 from NCATS/NIH), Vanderbilt University Medical Center institutional funding and Patient-Centered Outcomes Research Institute (PCORI; contract CDRN-1306-04869).